\documentclass[a4paper,12pt]{article}
 \pdfoutput=1
\usepackage[pdftex]{graphicx}
\usepackage{subfigure}

\newcommand{\be}{\begin{equation}}
\newcommand{\ee}{\end{equation}}

\newcommand{\bea}{\begin{eqnarray}}
\newcommand{\eea}{\end{eqnarray}}

\newcommand{\p}{\partial}

\newcommand{\nn}{\nonumber \\}
\newcommand{\f}{\frac}

\newcommand{\ra}{\rightarrow}
\begin{document}
\thispagestyle{empty}
\begin{flushright}
\end{flushright}
\begin{center} \noindent \Large \bf 
Weak Minimal Area In  Entanglement Entropy
\end{center}

\bigskip\bigskip\bigskip
\vskip 0.5cm
\begin{center}
{ \normalsize \bf   Shesansu Sekhar Pal${}^{a,b}$ and Shubhalaxmi Rath$^{b}$
}
\vskip 0.5 cm

{${}^a$Department of Physics, Utkal University, Bhubaneswar 751004, India}\\

{ ${}^b$Centre of Excellence in Theoretical and Mathematical Sciences\\
 Siksha \lq{}O\rq{} Anusandhan  University\\
 Khandagiri Square,  Bhubaneswar 751030, India}
\vskip 0.5 cm
\sf { shesansu@gmail.com}
\end{center}
\centerline{\bf \small Abstract}

We  re-visit  the minimal area condition of   Ryu-Takayanagi in the holographic calculation of  the  entanglement entropy.
In particular, the Legendre test and the Jacobi test. The necessary condition for the weak minimality  is checked via Legendre test and its sufficient nature via Jacobi test.  
We show  for AdS black hole with a strip type entangling region that  it is this minimality condition that makes the hypersurface not to cross the horizon, which is in  agreement with that  studied earlier  by   {\it Engelhardt et al.} and {\it Hubeny} using a different approach. 
Moreover, demanding the weak minimality condition on the  entanglement entropy  functional with  the higher derivative term  puts a constraint on the  Gauss-Bonnet  coupling: that is  there should be an upper bound on the value of the coupling, $\lambda_a< \f{(d-3)}{4(d-1)}$.

\newpage
\section{Introduction}

The recent conjecture on the holographic formulation  of the entanglement entropy by Ryu-Takayanagi (RT) \cite{Ryu:2006bv}   has given a new direction to do explicit calculations in the field theory provided  it admits a dual gravitational description\footnote{In a recent development in \cite{Bianchi:2012ev}, the authors have conjectured the existence of a geometric entropy in a theory of quantum gravity that includes it in  the entanglement entropy .} \cite{Maldacena:1997re}. 
In order to compute the entanglement entropy of a given region, $A$, with its complement in the field theory, it  proposes with a fixed time slice to consider a co-dimension two hypersurface, $\Sigma$, in the bulk in such a way that its boundary  coincides with the boundary of the region under study, i.e., $\p A=\p\Sigma$. Moreover, we  need to consider the  hypersurface that  minimizes the area. In which case,   the entanglement entropy is simply given by the area of the hypersurface divided by $4G_N$, where $G_N$ is the Newton\rq{}s constant and   it reads as
\be
S_{EE}(A)=Lim_{\p\Sigma=\p A}\f{{\rm Min}~(Area\left(\Sigma\right))}{4G^{d+1}_N}.
\ee

Recall that the area of a co-dimension two hypersurface is given by 
\be\label{area}
Area(\Sigma)=\int_{\Sigma} d^{d-1}\sigma \sqrt{det\left(\p_aX^M\p_b X^NG_{MN}\right)},\quad g_{ab}\equiv \p_aX^M\p_b X^NG_{MN},
\ee
where $X^M$ and $G_{MN}$ are the embedding functions and the bulk geometry, respectively. Setting the first variation of such an area functional to zero gives the following equation, which is essentially the equation of the hypersurface \cite{Hubeny:2007xt}, and is further studied\footnote{Some other interesting studies are reported in \cite{Alishahiha:2013dca}.} in \cite{Chen:2013qma, Pal:2013fha, Erdmenger:2014tba, Bhattacharyya:2014yga}
\be\label{eom_area_X}
g^{ab}{\cal K}^S_{ab}=0,\quad {\rm and}\quad {\cal K}^S_{ab}=\p_a\p_b X^S-\gamma^c_{ab}\p_c X^S+\p_aX^M\p_bX^N\Gamma^S_{MN},
 \ee 
where $g^{ab}$ is the inverse of the induced metric, $g_{ab}$. $\gamma^c_{ab}$ and $\Gamma^S_{MN}$ are the connections defined with respect to the induced metric on the hypersurface and the bulk geometry, respectively. 

In order to find the entanglement  entropy, we can solve for $X^M$\rq{}s in  eq(\ref{eom_area_X}) for a given bulk geometry and substitute that into the area integral. However, it is not ${\it a~ priori}$ clear  that 
the solution of  eq(\ref{eom_area_X}) will  necessarily give us a minimum area. It can give a maximum, a minimum or a  point of inflection/saddle point.
It is suggested in  \cite{Myers:2012ed} that by working with  {\it the Euclidean signature, the extremization of the area functional will automatically give a global minimum of the area functional. However, with the Minkowski signature, the extremization gives saddle points and one need to opt for the solution that gives a minimum area}.  

In this paper we want to study the (weak) minimal  condition on the entanglement entropy functional  with the Minkowski signature  for generic $\Sigma$ that follows from eq(\ref{area}) and  study the consequences through some examples.

In order to check the minimality condition on the area or equivalently on the entanglement entropy functional, let us find the second variation of the area functional eq(\ref{area}), which gives
\bea\label{sec_var}
\delta^2 Area(\Sigma)&=&\int \sqrt{det~g_{ab}}\Bigg[\left((g^{ab}g^{cd}-2g^{ac}g^{bd})G_{KL}G_{MN}\p_bX^N\p_dX^L+g^{ac}G_{MK}\right)
\p_c\delta X^K\p_a\delta X^M\nn&&+\left((g^{ab}g^{cd}-2g^{ac}g^{bd})G_{KL}\p_PG_{MN}\p_d X^L\p_a X^M\p_b X^N
+2g^{bc}\p_b X^N\p_PG_{KN} \right)\nn&&\p_c\delta X^K \delta X^P +\bigg(\f{1}{4}(g^{ab}g^{cd}-2g^{ac}g^{bd}) \p_a X^M\p_b X^N \p_PG_{MN}\p_c X^S\p_d X^L \p_K G_{SL}+\nn&&\f{1}{2}g^{ab}\p_a X^M\p_b X^N \p_P\p_KG_{MN} \bigg)\delta X^P\delta X^K\Bigg]\nn
&=&\int V^T\cdot M \cdot V,
\eea
where the column vector  $V=
\left(
    \begin{array}{c}
      \p~ \delta X \\
      \delta X
    \end{array}
  \right)$
and we have dropped the indices, for simplicity. 
Note that in getting the result, we have dropped a total derivative term,  which essentially will give  a boundary term and we assume that it is not going to contribute at the boundary. Also, a term proportional to the equation of motion. If we want the area to be a minimum then the determinant of the matrix $M$  should be positive. The Jacobi test says about the positivity of the matrix $M$ and it corresponds to the sufficient condition for the weak minimum.

In calculus the Legendre test says that 
\be\label{min_area}
\f{\delta^2 (Area(\Sigma))}{\p_c\delta X^K\p_a\delta X^M}=2\sqrt{det~(g_{ab})}\left[(g^{ab}g^{cd}-2g^{ac}g^{bd})G_{KL}G_{MN}\p_bX^N\p_dX^L+g^{ac}G_{MK}\right]> 0
\ee
 and it gives a weak condition on the minimality of the function, in this case the area. Generically, it is very difficult to combine eq(\ref{eom_area_X}) and eq(\ref{min_area}) so as to draw any useful conclusion\footnote{ However, it is certainly very interesting to find  connection between eq(\ref{min_area}) and  with the extrinsic curvature as proposed in the context of black holes  in \cite{Engelhardt:2013tra}, if any.}. Instead, 
 in what follows, we shall calculate the quantity, eq(\ref{min_area}),  in different examples and check whether the area is (weak) minimum or not.

In this paper we  study the consequence of  such weak minimality condition in different spacetime, such as AdS spacetime with and without the  black holes, hyperscale violating geometries and geometries with higher derivative terms.  In the case of the black hole geometry, the minimal area condition of 
 the RT conjecture gives us a very  interesting  consequence  that is the spacelike hypersurfaces do not cross the horizon. This   conclusion matches precisely as studied  in \cite{Hubeny:2012ry}, where the author did not find any solution to the embedding field, $X^M$, of eq(\ref{eom_area_X}) inside the horizon and further studied in \cite{Pal:2013fha} at finite \lq{}t Hooft coupling
and more generally in \cite{Engelhardt:2013tra}.  

By studying different  examples, we find that the second variation of the area functional can be written as
\be
\delta^2 Area(\Sigma)=\int \prod_i dx_i \left( A ~\delta r\rq{}\delta r\rq{}+B \delta r\delta r\rq{}+C~\delta r \delta r \right),
\ee
where $r\rq{}=\f{dr}{dx}$ and $x$ is one of the coordinate on the hypersurface.
The integral is over the world volume coordinates of the co-dimension two hypersurface. 

The weak mimimality condition states that  the second variation of the area functional with respect to $r'$ becomes positive
\be
A>0,
\ee
and the Jacobi test says  the determinant of the matrix $M$ should be positive,  $AC-\f{B^2}{4}>0$.
In this paper, we shall be checking these conditions by studying several examples. 

It is also very interesting to ask the  minimal nature  of the entanglement entropy functional even in the finite \lq{}tHooft coupling limit\footnote{A prescription is given in   \cite{Dong:2013qoa,Camps:2013zua} to construct the entanglement entropy functional in such cases.}. In this context, it is argued in \cite{Fursaev:2013fta} and \cite{Dong:2013qoa} based on the strong subadditivity property  that the first possible higher derivative correction to the entanglement entropy functional indeed obeys the minimality condition. 
For our purpose, we consider the following  entanglement entropy functional,  as also studied in \cite{ Fursaev:2006ih, deBoer:2011wk} and \cite{Hung:2011xb}
\be
4 G_N S_{EE}=\int d^{d-1}\sigma~\sqrt{det( g_{ab})}\left[1+\f{2\lambda_a R^2_A}{(d-2)(d-3)} R(g)\right],
\ee
where $R(g)$ denotes the Ricci scalar made out of the induced metric. We found the following constraint upon demanding the weak minimality of the entanglement entropy functional
\be
\lambda_a< \f{(d-3)}{4(d-1)}.
\ee

Note that we denote $R_A$ as the radius of the AdS spacetime. The constraint  on the Gauss-Bonnet coupling, $\lambda_a$, does not coincide with the result obtained in \cite{Buchel:2009sk,Camanho:2009hu}. So it means the hypersurface under study does not have either 
minimal or maximal entanglement entropy. The maximal  entanglement entropy is ruled out otherwise the Gauss-Bonnet coupling can be as large as infinity. Through this study there follows an important result that is theories without higher derivative terms does admit a minimal hypersurface but not with\footnote{The caveat is that the weak minimality analysis is performed only to leading order in the coupling.}. Hence, the nature of the hypersurface with the higher derivative term remains to be seen in future.

\section{Example: Strip type} 

In this section, we shall check  the minimality of the area functional by doing some explicit calculation for the strip type entangling region. This will be performed by finding  the embedding field that follows from eq(\ref{eom_area_X}). The strip on the field theory is defined as $0\leq x_1\leq \ell$ and $-L/2\leq (x_2,\cdots,x_{d-1})\leq L/2$. Moreover, the bulk  geometry is assumed to take  the following form
\be
ds^2_{d+1}=-g_{tt}(r)dt^2+g_{xx}(r)(dx^2_1+\cdots+dx^2_{d-1})+g_{rr}(r)dr^2.
\ee

With the following embedding fields $X^t=0,\quad X^a=x^a=\sigma^a,\quad X^r=r(x_1)$, the induced metric is
\be\label{ind_metric_strip}
ds^2_{d-1}=g_{ab}d\sigma^ad\sigma^b=g_{xx}(r)(dx^2_2+\cdots+dx^2_{d-1})+\left(g_{rr}(r)r\rq{}^2+g_{xx}(r)\right)dx^2_1,\quad r\rq{}\equiv \f{dr}{dx_1}.
 \ee 

In this case, the area takes the following form: $Area=L^{d-2}\int dx_1 g^{\f{d-2}{2}}_{xx}\sqrt{g_{xx}+g_{rr}r\rq{}^2}$, whose
second variation  gives the following column vector, $V$, and the matrix, $M$ 
\be
M=\left(
    \begin{array}{c l}
     A & \f{B}{2} \\
     \f{ B}{2 }& C
    \end{array}
  \right),\quad 
V= \left(
    \begin{array}{c }
      \delta r\rq{}\\
      \delta r
    \end{array}
  \right).
\ee

This means 
\be\label{2nd_var_area}
\delta^2 Area(\Sigma)=L^{d-2}\int \left(A ~\delta r\rq{}\delta r\rq{}+B~ \delta r \delta r\rq{}+C~\delta r \delta r \right)=L^{d-2}\int A\left( \delta r\rq{}+\f{B}{2A}\delta r\right)^2+\f{(4AC-B^2)}{4A}\delta r\delta r.
\ee

In order to have a minimum area functional  $A$ should be positive and $4AC> B^2$. 
Note, the determinant of the matrix $M$ is  $det(M)=AC-\f{B^2}{4}$ and $\f{\delta^2 (Area(\Sigma))}{\delta r\rq{} \delta r\rq{}}\sim2 A$, where the expressions for these quantities are
\bea\label{a_b_c}
A&=&\f{g^{d/2}_{xx}g_{rr}}{(g_{xx}+r\rq{}^2g_{rr})^{3/2}},\quad
B=g^{\f{d-2}{2}}_{xx} r\rq{}\left(\f{(d-2)g\rq{}_{xx}g_{rr}+2g\rq{}_{rr}}{g_{xx}\sqrt{g_{xx}+r\rq{}^2g_{rr}}}\right)-\f{g^{\f{d-2}{2}}_{xx}g_{rr}r\rq{}(g\rq{}_{xx}+r\rq{}^2g\rq{}_{rr})}{(g_{xx}+r\rq{}^2g_{rr})^{3/2}},\nn
C&=&-\f{g^{\f{d-2}{2}}_{xx}(g\rq{}_{xx}+r\rq{}^2g\rq{}_{rr})^2}{4(g_{xx}+r\rq{}^2g_{rr})^{3/2}}+\left(\f{d-2}{4}\right) g^{\f{d-6}{2}}_{xx}\sqrt{g_{xx}+r\rq{}^2g_{rr}}\left((g\rq{}^2_{xx}(d-4)+2g_{xx}g\rq{}\rq{}_{xx}\right)+\nn
&&\f{g^{\f{d-2}{2}}_{xx}(g\rq{}\rq{}_{xx}+r\rq{}^2g\rq{}\rq{}_{rr})+(d-2)g^{\f{d-4}{2}}_{xx}g\rq{}_{xx}(g\rq{}_{xx}+r\rq{}^2g\rq{}_{rr})}{2\sqrt{g_{xx}+r\rq{}^2g_{rr}}}.
\eea

The meaning of the derivative is as follows: $g'_{ab}\equiv\f{\p g_{ab}}{\p r}$ and $r'\equiv\f{d r}{dx_1}$.
Generically, it is very difficult to draw any conclusion on the determinant of matrix $M$. However, it is easy to show that the quantity $A$  is  positive. This follows by considering the solution that follows, in fact  as constructed in \cite{Pal:2013fha}, $r\rq{}=\f{\sqrt{g^d_{xx}(r)-g^{d-1}_{xx}(r_{\star})g_{xx}(r)}}{g^{\f{d-1}{2}}_{xx}(r_{\star})\sqrt{g_{rr}(r)}}$, in which case
\be\label{exp_A}
A=\f{g_{rr}(r)g^{\f{3(d-1)}{2}}_{xx}(r_{\star})}{g^d_{xx}(r)}> 0,
\ee
and the  expression for $det(M)$ are very cumbersome to write  down explicitly. The quantity, $r_{\star}$, is determined by requiring that $r\rq{}$ vanishes there.

Note, $\f{1}{L^{d-2}}\f{\delta^2 (Area(\Sigma))}{\delta r\rq{} \delta r\rq{}}=2 A$. 
In order to check the weak minimality  condition on the area functional, we need to look at the condition  $A > 0$, which is obeyed automatically. Now moving onto determine the sign of the  determinant of the matrix $M$,
generically, it is very difficult to draw any conclusion. Nevertheless, 
 we shall check it   on case-by-case basis. 

\paragraph{AdS:} To begin with, let us consider the AdS spacetime with radius $R$ and  the boundary is at $r=0$, in which case
\be
A=r^{2(d-1)}R^{d-1} r^{-3(d-1)}_{\star},\quad B=-\f{2(d-1)}{r}R^{d-1}\sqrt{r^{2-2d}-r^{2-2d}_{\star}},\quad C=\f{d(d-1)}{r^{2d}}\left(Rr_{\star}\right)^{d-1},
\ee
where we have considered $g_{xx}=R^2/r^2=g_{rr} $. The quantity 
\be
\f{4 A C-B^2}{4}=(d-1)^2\f{R^{2(d-1)}}{r^{2d}}\left[\f{(2d-1)}{(d-1)}\left(\f{r}{r_{\star}}\right)^{2(d-1)}-1 \right].
\ee
We know that the surface under study starts from the boundary $r=0$ and goes all the way to $r=r_{\star}$ but does not go past $r=r_{\star}$, which means the above quantity is positive only close to $r_{\star}$, whereas close to UV, it becomes negative. This result
suggests that the weak minimum   is not a sufficient condition. 

\paragraph{HSV:} For Hyperscale violating (HSV) solution in the convention  of \cite{Pal:2012zn} with $g_{xx}=R^2/r^{2-2 \gamma}=g_{rr}$ where $\gamma $ is a constant. The positivity of $A$ is easy to observe whereas  the $det(M)$ is

\be
\f{4 A C-B^2}{4}=-(d-1)^2(\gamma-1)^2\f{R^{2(d-1)}}{r^{2d-2\gamma(d-1)}}\left[\f{(2d-1-2\gamma(d-1) )}{(\gamma-1)(d-1)}\left(\f{r}{r_{\star}}\right)^{2(d-1)(1-\gamma)}+1\right].
\ee

It is easy to see again that close to UV, the $det(M)$ becomes negative and becomes positive close to $r_{\star}$ for both positive and negative $\gamma$.

\paragraph{Black hole: }
Let us consider a black hole, for simplicity, we assume it asymptotes to  AdS spacetime with the boundary to be at $r=0$.  In this coordinate system the horizon is located at $r=r_h> 0$. Moreover, $g_{xx}(r)$ is positive for all values of $r$ and it takes the following form

\[ g_{rr}(r) = \left\{
  \begin{array}{l l}
    +ve & \quad \textrm{for $r< r_h$ \quad (Outside the horizon)}\\
    -ve & \quad \textrm{for $r> r_h$\quad    (Inside the horizon)}.
  \end{array} \right. \]
 
 It follows from eq(\ref{exp_A}) that as the hypersurface goes inside the black hole, the quantity, $A$, becomes negative whereas outside the horizon, it stays positive.  So, we see that if the hypersurface  stays outside the horizon, as suggested in \cite{Engelhardt:2013tra} and \cite{Hubeny:2012ry}, then it follows naturally that there exists a   (weak) minimality  condition on the area functional.



In order to check the sign of the determinant of the matrix $M$, 
let us take the following choice of the metric components
\be
g_{xx}=R^2/r^2,\quad g_{rr}=R^2/(r^2 f(r)),\quad f(r)=1-(r/r_h)^d,
 \ee 
 In which case we get
 \bea
A&=&R^{d-1}\f{r^{2(d-1)}}{f(r)}r^{-3(d-1)}_{\star},\nn  B&=&R^{d-1}r^{d/2}_h\f{\sqrt{r^{2(1-d)}-r^{2(1-d)}_{\star}}}{r^3r^{2 d}_{\star}(r^d_h-r^d)^{3/2}}\left(d[r^{3d}r^2_{\star}+r^{2d}_{\star}(3r^{d+2}-2r^2r^d_h)]-2r^2r^{2d}_{\star}(r^d-r^d_h) \right),\nn
C&=&\f{dR^{d-1}}{4(r^d-r^d_h)^2}r^{2(d-2)}r^{d-1}_{\star}\Bigg((9d-6)r^{2(2-d)}-10(d-1)r^{4-3d}r^d_h+4(d-1)r^{4(1-d)}r^{2d}_h -\nn&&dr^{2d}r^{4(1-d)}_{\star}-2(2d-1)r^2 r^{2(1-d)}_{\star}+2(d-1)r^{2-d}r^d_hr^{2(1-d)}_{\star}\Bigg)
 \eea
 where $r_{\star}$ is the turning point of the solution, which  is the maximum reach of the hypersurface in the bulk. 
 
 Let us re-scale: $r=u r_{\star}$ and $r_h= n r_{\star}$, so that $u$ and $n$ are dimensionless. For simplicity, we take $d=4$, in which case
 \be
  4AC-B^2=-4n^4R^6\f{(3n^8(7u^6-3) -6u^4n^4(u^{12}-5+8u^6)+u^8(2u^{12}-25+35u^6))}{r^8_{\star}u^8(u^4-n^4 )^3}
 \ee
Generically, $\f{r^8_{\star}}{R^6} \left(4AC-B^2\right)$ is a function of two variables $n$ and $u$. It is very easy to see that 
close to UV i.e., for very small values of $u$, the function  $\f{r^8_{\star}}{R^6} \left(4AC-B^2\right)$ becomes negative. It means close to the boundary the determinant of matrix $M$ is not positive.  So the weak minimality condition is not  a sufficient condition. 

Let us recall that $r=\left(\f{u}{n}\right)r_h$. It means when $u> n$ we are inside the horizon and for $u< n$  outside the horizon. If the turning point $r_{\star}$ is inside the horizon then $r=r_{\star}>r_h$. This means $n<1$. Similarly, for  $r_{\star}$  outside the horizon then $r=r_{\star}<r_h$, which means $n>1$. In summary

\[  \left\{
  \begin{array}{l l }
    u<n, & n>1 \quad \textrm{\quad (Outside the horizon)}\\
    u>n,  & n<1 \quad \textrm{\quad    (Inside the horizon)}
  \end{array} \right.\]
For simplicity, we shall restrict $n$ to stay from $1< n\leq  2$ for outside the horizon which means $0\leq u < 1$.  Whereas for inside the horizon, we shall take $0\leq n < 1$ means $1 < u \leq 2$. 

 \begin{figure}[t]\label{Fig}
   {\includegraphics[ width=8cm,height=6cm]{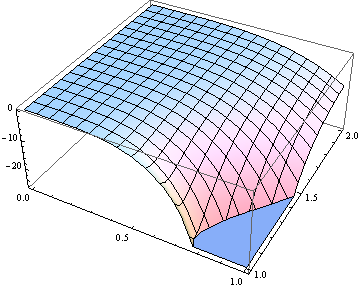} }
  \caption{
  $\f{r^8_{\star}}{R^6} \left(4AC-B^2\right)$ is plotted for $AdS_5$ black hole  inside the horizon  for which $0\leq n < 1$ and $1 < u \leq 2$.  }
\end{figure}

The quantity $\f{r^8_{\star}}{R^6} \left(4AC-B^2\right)$ is plotted  inside the horizon for AdS black hole in $4+1$ dimensional spacetime  in fig(1). It is easy to notice that this quantity is always negative inside the horizon of the  AdS black hole. 

Both the quantities, $A$, and the determinant of the matrix $M$ becomes negative. 
This simply means there does not exists any hypersurface inside the horizon that minimizes the area functional. Recall, according to  
Ryu-Takayanagi  conjecture, we need to find the area of the hypersurface that minimizes the area functional in the computation of the entanglement entropy. 
So, we can interpret the absence of the minimal area  hypersurface inside the horizon as the non-penetration of such  hypersurface into the horizon. This conclusion is in perfect agreement with that reached  in \cite{Engelhardt:2013tra} and \cite{Hubeny:2012ry}.

\paragraph{Outside the horizon:}

\begin{figure}[t]\label{fig_2}
   {\includegraphics[ width=8cm,height=6cm]{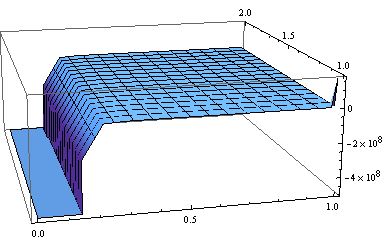} }
  \caption{
   The  figure is  is plotted for $\f{r^8_{\star}}{R^6} \left(4AC-B^2\right)$r $AdS_5$ black hole  outside the horizon  for which $0\leq u < 1$ and $1 < n \leq 2$.   }
\end{figure}

Let us look at the  behavior of the quantities $A$ and $det(M)$ outside the horizon. 
It is easy to see that the quantity $A$ is always positive outside the horizon, which follows simply from  eq(\ref{exp_A}). The information about the other quantity, namely, the 
determinant of  the matrix $M$ can be obtained numerically, which is plotted in fig(\ref{fig_2}).

It is clear from  figure fig(\ref{fig_2}) that the determinant of  the matrix $M$ becomes negative close to UV, which suggests that the Jacobi condition for the sufficient nature  of the weak minimality condition does not hold.

\paragraph{Confining Solution:}

Let us study the weak minimality condition on the area functional in the case for which the background solution shows confining behavior. To generate such a confining background, the easiest method is to start with the uncharged black hole solution and perform a double Wick rotation. In the end the solution that asymptotes to  $AdS_{d+1}$ with unit AdS radius reads as 
\be
ds^2_{d+1}=\f{1}{r^2}\left(-dt^2+f(r) ~dx^2_1+dx^2_2+\cdots+dx^2_{d-1}\right)+\f{dr^2}{r^2f(r)},\quad f(r)=1-\f{r^d}{r^d_0}.
\ee
The coordinate $x_1$ is now periodic with periodicity $2\pi\beta$, whose explicit form is not important for us. The IR is at  $r=r_0$ and the UV is at $r=0$.
We can  proceed further by studying two cases, depending on the fields that we are exciting. 

\paragraph{Case 1:}The induced metric on the co-dimension two hypersurface takes the following form
\be
ds^2_{d-1}=\f{1}{r^2}\left(f~dx^2_1+dx^2_3+\cdots+dx^2_{d-1}\right)+\left(1+\f{r\rq{}^2}{f}\right)\f{dx^2_2}{r^2},\quad r\rq{}=\f{dr}{dx_2}
\ee

In which case, the area of the induced geometry for the strip times a a shrinking circle type entangling region, $0\leq x_1\leq 2\pi \beta,~ 0\leq x_2\leq \ell,~-L/2 \leq (x_3,\cdots,x_{d-1})\leq L/2$,  becomes
\be
A=\int dx_1\cdots dx_{d-1}~ \f{\sqrt{f(r)+r\rq{}^2}}{r^{d-1}}.
\ee

The solution to the equation of motion  takes the following form
\be
\f{dr}{dx_2}=\f{\sqrt{f(f-c^2_0 r^{2(d-1)})}}{c_0 r^{d-1}},
\ee
where the constant of integration $c_0$ is determined as follows: $\left( \f{dr}{dx_2}\right)_{r_{\star}}\ra 0$. This means $c_0=\f{\sqrt{f(r_{\star})}}{r^{d-1}_{\star}}$. The second variation of the area functional can be written as follows 
\be
\delta^2 Area=\int \prod_i dx_i \left(A~ \delta r\rq{}\delta r\rq{}+{\widetilde A}~ \delta r\delta r\right),
\ee
where we have dropped a boundary term using the boundary condition $\delta r(0)=0$ and $\delta r(\ell)=0$. The quantities are 
\bea
A&=&\f{1}{r^{d-1}\left(f(r)+r\rq{}^2\right)^{3/2}},\nn
{\widetilde A}&=&\f{d(d-1)}{r^{d+1}}\sqrt{f(r)+r\rq{}^2}-\f{(d-1)}{r^d\sqrt{f(r)+r\rq{}^2}}\f{df}{dr}+\f{d}{dx_2}\left(\f{(d-1)r\rq{}}{r^d\sqrt{f(r)+r\rq{}^2}} \right)-\nn &&\f{1}{4r^{d-1}(f+r\rq{}^2)^{3/2}}\left(\f{df}{dr}\right)^2+\f{d}{dx_2}\left(\f{r\rq{} (df/dr)}{2r^{d-1}(f(r)+r\rq{}^2)^{3/2}}\right)+\f{1}{2r^{d-1}\sqrt{f(r)+r\rq{}^2}} \f{d^2}{dr^2}f.\nn
\eea

Once again we can introduce the function $W$ as is done in the introduction and finally we are interested in the quantity $A$.  Using the solution for $r\rq{}$ results in
\be
A=\f{r^{2(d-1)}c^3_0}{(f(r))^3}>0.
\ee

It is easy to see the positivity of, $A$, because the radial coordinate stays from $0\leq r\leq r_0$. Hence, the weak minimality of the area functional for this case is checked.

\paragraph{Case 2:} In this case, we consider the embedding field as studied in \cite{Klebanov:2007ws}, i.e., the field, $r$, that is excited is a function of the compact coordinate $x_1$. In the $(r,~x_1)$ plane it will be a cigar. In which case the induced geometry reads as

\be
ds^2_{d-1}=\f{1}{r^2}\left(dx^2_2+dx^2_3+\cdots+dx^2_{d-1}\right)+\left(\f{f^2+r\rq{}^2}{r^2f}\right)+dx^2_1,\quad r\rq{}=\f{dr}{dx_1}
\ee

The area functional reads as
\be
Area=\int dx_1\cdots dx_{d-1}~\left(\f{\sqrt{f^2+r\rq{}^2}}{r^{d-1}\sqrt{f}}\right).
\ee

The equation of motion that follows gives the following solution
\be
r\rq{}=\f{f(r)\sqrt{f(r)-c^2_0r^{2(d-1)}}}{c_0 r^{d-1}},
\ee
where the constant of integration, $c_0$, is found by demanding that the quantity, $r\rq{}$, vanishes in the limit $r\ra r_{\star}$. It sets $c_0=\f{\sqrt{f(r_{\star})}}{r^{d-1}_{\star}}$. On finding the second variational of the area functional using the boundary condition $\delta r(0)=0$ and $\delta r(2\pi \beta)=0$ gives 
\be
\delta^2 Area=\int \prod_i dx_i \left(A~ \delta r\rq{}\delta r\rq{}+{\widetilde A}~ \delta r\delta r\right).
\ee
For our purpose, the precise form of the quantity ${\widetilde A}$ is not important  as we are interested to find only the form of $A$ and its sign.  In the present case, it reads as
\be
A=\f{1}{r^{d-1}\left(f(r)+\f{r\rq{}^2}{f(r)}\right)^{3/2}}.
\ee
Using the solution as written above, it is easy to conclude that
\be
A=\f{r^{2(d-1)}c^3_0}{(f(r))^3}>0.
\ee

It is interesting to note that the quantity, $A$, in both the cases, $r(x_1)$ and $r(x_2)$ gives a minimum to the area functional. 

As an aside, the existence of two valid configurations
 means that there can be a phase transition induced quantum mechanically depending on the energy of these two configurations, which is  studied in detail in  \cite{Klebanov:2007ws}. But for our purposes we see that both are becoming minima to the area functional, which we set out to find.

\subsection{Sphere}

Let us consider another example, where the entangling region, $\Sigma$,  is of the sphere type. In this context, we assume that the bulk geometry is
\be
ds^2_{d+1}=-g_{tt}(r)dt^2+g_{xx}(r)(dx^2_1+\cdots+dx^2_{d-1})+g_{rr}(r)dr^2
=-g_{tt}dt^2+g_{xx}(d\rho^2+\rho^2d\Omega^2_{d-2})+g_{rr}dr^2
\ee
Using the Ryu-Takayanagi prescription, the geometry of the co-dimension two hypersurface takes the following form
\be
ds^2_{d-1}=(g_{xx}+g_{rr}r\rq{}^2)d\rho^2+g_{xx}\rho^2 d\Omega^2_{d-2},
\ee
where $r\rq{}=\f{dr}{d\rho}$. The area  functional  reads  as
\be\label{area_sphere}
Area(\Sigma)=\omega_{d-2}\int d\rho \rho^{d-2} g^{\f{d-2}{2}}_{xx}\sqrt{g_{xx}+g_{rr}r\rq{}^2}=\omega_{d-2}\int dr \rho^{d-2} g^{\f{d-2}{2}}_{xx}\sqrt{g_{xx}\rho\rq{}^2+g_{rr}},
\ee
where $\omega_{d-2}$ is the volume form associated to the unit $d-2$ dimensional sphere, $S^{d-2}$.
The equation of motion that follows takes the following form 
\be
\p_r\left(\f{\rho^{d-2}g^{d/2}_{xx}\rho\rq{}}{\sqrt{g_{xx}\rho\rq{}^2+g_{rr}}} \right)-(d-2) \rho^{d-3} g^{\f{d-2}{2}}_{xx}\sqrt{g_{xx}\rho\rq{}^2+g_{rr}}=0,
\ee
where $\rho\rq{}=\f{d\rho}{dr}$. Upon considering the  background geometry as AdS spacetime with radius $R$, $g_{xx}=g_{rr}=R^2/r^2$,  the solution that follows takes the following form:  
$\rho=\sqrt{c^2-r^2}$, where $c$ is a constant of integration.

Let us find the second variation of the area functional as written in eq(\ref{area_sphere}) for AdS soacetime 
\be
\delta^2 Area(\Sigma)=\omega_{d-2}R^{d-1}\int d\rho \left[ A(\delta r\rq{})^2+B \delta r\delta r\rq{}+C (\delta r)^2\right],
\ee
where 
\be
A=\f{\rho^{d-2}}{r^{d-1}(1+r\rq{}^2)^{3/2}},\quad B=-2(d-1)\f{r\rq{}\rho^{d-2}}{r^d\sqrt{1+r\rq{}^2}},\quad C=d(d-1)\f{\rho^{d-2}\sqrt{1+r\rq{}^2}}{r^{d+1}}.
\ee
 In getting the above mentioned second variation of the area functional, we have used the equation of motion obeyed by $r=r(\rho)$. 
On computing  the following quantity
\be
\f{\delta^2 (Area(\Sigma))}{\delta r\rq{} \delta r\rq{}}=2 \f{\omega_{d-2}g^{d/2}_{xx}g_{rr}\rho^{d-2}}{(g_{xx}+r\rq{}^2g_{rr})^{3/2}}= 2 \f{\omega_{d-2}R^{d-1}\rho^{d-2}}{c^3 (c^2-\rho^2)^{\f{d-4}{2}}}=2 \f{\omega_{d-2}R^{d-1}\rho^{d-2}}{c^3 r^{d-4}}>0.
\ee

In getting the second equality, we have used the geometry of AdS spacetime. Note that both $r$ and $\rho$ are real and positive, hence the above quantity, $A$, is positive. 

Let us determine the sign associated to the determinant of matrix $M$. In which case
\be
4AC-B^2=-4(d-1)\rho^{2(d-2)}\left(\f{(d-1)r'^2-d}{r^{2d}(1+r'^2)}\right)=4(d-1)\rho^{2(d-2)}\left(\f{\rho^2+d(r^2-\rho^2)}{c^2r^{d/2}}\right)
\ee
Using the 
following hypersurface:  $r=\sqrt{c^2-\rho^2}$ close to UV, we find the quantity, $4AC-B^2$, close to UV  
becomes negative and large, which  means the determinant of the matrix $M$ is negative.
Note that the constant $c$ can be identified with the  size of the sphere, $R$.

\section{With higher derivative}

In the presence of the higher derivative term in the entanglement entropy functional,  it is not {\it a priori} clear  that the   entanglement entropy functional will be a minimum, automatically. 
Moreover, we cannot apply eq(\ref{min_area}) in the determination of the Legendre test. However, it is suggested in \cite{Fursaev:2013fta} and \cite{ Dong:2013qoa} that for a very specific type of entanglement entropy functional one can get a minimal entanglement entropy functional.  In the present case, we shall determine the consequence of  the imposition of the minimal nature of the entanglement entropy functional  for the AdS spacetime only, which depends crucially  on the value of the coupling, $\lambda_1$, as defined latter. The precise form of the entanglement entropy functional 
with the higher derivative term  can be considered as described  by the Jacobson-Myers functional \cite{Jacobson:1993xs}. In fact, for our purpose, we  shall consider the  structure as studied  in \cite{Hung:2011xb, Fursaev:2006ih} and \cite{deBoer:2011wk}
 \be\label{assumption_EE}
4 G_N S_{EE}=\int d^{d-1}\sigma~\sqrt{det( g_{ab})}\left[1+\lambda_1 R(g)\right].
 \ee
where $\lambda_1$  is the coupling constant and defined\footnote{This follows by comparing with the action used in \cite{Hung:2011xb}.} as $\lambda_1\equiv\f{2\lambda_a R^2_A}{(d-2)(d-3)}$.
Let us evaluate the entanglement entropy for the strip type entangling region as discussed earlier. Using the structure of the  induced metric $g_{ab}$ as written down in eq(\ref{ind_metric_strip}) gives \cite{Pal:2013fha}
  \bea\label{ee_strip_lambda1}
2 G_N   &S_{EE}&=L^{d-2}\int dr \f{g^{\f{d-6}{2}}_{xx}}{4\left[g_{rr}+g_{xx}x\rq{}^2_1\right]^{\f{3}{2}}}\Bigg[4g^2_{xx}(g_{rr}+g_{xx}x\rq{}^2_1)^2+\lambda_1(d-2) \bigg(2g_{xx}g\rq{}_{xx}g\rq{}_{rr}-\nn&&
   (d-7)x\rq{}^2_1g_{xx}g\rq{}^2_{xx}+4x\rq{}_1x\rq{}\rq{}_1g^2_{xx}g\rq{}_{xx}-4x\rq{}^2_1 g^2_{xx}g\rq{}\rq{}_{xx}-4 g_{xx}g_{rr}g\rq{}\rq{}_{xx}-(d-5)g_{rr}g\rq{}^2_{xx} \bigg)\Bigg]\nn
   \eea
where $x\rq{}_1=\f{dx_1}{dr}$. This for the  AdS spacetime\footnote{Such solutions and the associated phase transitions with the black hole solutions are studied in great detail in e.g., \cite{Cai:2001dz}, \cite{Nojiri:2001aj} and \cite{Cho:2002hq}.} with   the boundary at $r=0$ and with the AdS  radius $R_0$ becomes  
\bea
2G_N S_{EE}&=&L^{d-2}R^{d-1}_0 \int dx_1\Bigg[ \f{\sqrt{1+r\rq{}^2}}{r^{d-1}}-\f{(d-2)\lambda_1}{R^2_0}\left(\f{(d-1)r\rq{}^2}{r^{d-1}\sqrt{1+r\rq{}^2}} -\f{2r\rq{}\rq{}}{r^{d-2}(1+r\rq{}^2)^{3/2}}\right) \Bigg]\nn&=&
L^{d-2}R^{d-1}_0 \int dr \Bigg[ \f{\sqrt{1+x\rq{}^2_1}}{r^{d-1}}-\f{(d-2)\lambda_1}{R^2_0}\left(\f{(d-1)(1+x\rq{}^2_1)+2rx\rq{}_1x\rq{}\rq{}_1}{r^{d-1}(1+x\rq{}^2_1)^{3/2}}\right) \Bigg].
\eea

The equation of motion that follows takes the following form 
\be\label{eom_higher_der}
\f{d}{dr}\left[\f{x\rq{}_1}{r^{d-1}\sqrt{1+x\rq{}^2_1}}-\f{(d-2)(d-3)\lambda_1 x\rq{}_1}{R^2_0r^{d-1}(1+x\rq{}^2_1)^{3/2}} \right]=0.
\ee

The second variation of the entanglement entropy functional can be expressed as

\bea
2 G_N   \delta^2S_{EE}&=&L^{d-2}R^{d-1}_0\int dx_1 \left(A~ \delta r\rq{}\delta r\rq{}+B\delta r\delta r\rq{}+C\delta r\delta r+D\delta r\delta r\rq{}\rq{}+E \delta r\rq{}\delta r\rq{}\rq{}\right)\nn&=&
L^{d-2}R^{d-1}_0\int dx_1 \quad V^T.~ M.~V,
\eea
where the column vector  $V$ and  the matrix $M$ are
\be
V=
\left(
    \begin{array}{c}
       \delta r' \\
      \delta r\\
       \delta r''
    \end{array}
  \right),\quad 
M=\left(
    \begin{array}{c lc}
     A & \f{B}{2}& \f{E}{2} \\
     \f{ B}{2 }& C &\f{D}{2}\\
      \f{E}{2}&\f{D}{2}& 0
    \end{array}
  \right),\quad {\rm det(M)}=\f{E(BD-CE)-AD^2}{4}.
\ee

The various expressions are
\bea
A&=&\f{1}{r^{d-1}(1+r\rq{}^2)^{3/2}}-\f{(d-2)\lambda_1}{R^2_0}\Bigg(\f{2(d-1)-3(d-1)r\rq{}^2}{r^{d-1}(1+r\rq{}^2)^{3/2}} +\f{3(d-1)r'^4}{r^{d-1}(1+r\rq{}^2)^{5/2}}+\nn&&
\f{6r''}{r^{d-2}(1+r\rq{}^2)^{5/2}}-\f{30r'^2r''}{r^{d-2}(1+r\rq{}^2)^{7/2}}\Bigg),\nn
B&=&-\f{2(d-1)r'}{r^{d}\sqrt{1+r\rq{}^2}}-\f{(d-2)\lambda_1}{R^2_0}\Bigg( -\f{4(d-1)^2r'}{r^d\sqrt{1+r'^2}}+\f{2(d-1)^2r'^3}{r^d(1+r'^2)^{3/2}}-\f{12(d-2)r'r''}{r^{d-1}(1+r'^2)^{5/2}}\Bigg),\nn
C&=&\f{d(d-1)\sqrt{1+r'^2}}{r^{d+1}}-\f{(d-2)\lambda_1}{R^2_0}\Bigg(\f{d(d-1)^2r'^2}{r^{d+1}\sqrt{1+r'^2}}-\f{2(d-1)(d-2)r''}{r^d(1+r'^2)^{3/2}}\Bigg),\nn
D&=&-\f{(d-2)\lambda_1}{R^2_0} \f{4(d-2)}{r^{d-1}(1+r'^2)^{3/2}},\quad E=-\f{(d-2)\lambda_1}{R^2_0} \f{12r'}{r^{d-2}(1+r'^2)^{5/2}}.
\eea 

Now, we can demand  the Legendre condition as stated earlier and at the end, we are interested to determine under what condition  the quantity $A$ is positive?  Using the real valued solution that follows from eq(\ref{eom_higher_der}) to the leading order in the coupling $\lambda_1$ gives
\be
A=r^{2(d-1)}r^{-3(d-1)}_{\star}\left(1-2(d-1)(d-2)\f{\lambda_1}{R^2_A} \right)+{\cal O}(\lambda_1)^2,
\ee
where we have used the relationship between the sizes of the AdS spacetime, $R_0$ and  $R_A$. The size $R_A$ is  defined in the infinite \lq{}tHooft coupling limit and is related as $R_0=R_A/\sqrt{f_{\infty}}$, where $f_{\infty}$ obeys the following relation: $1-f_{\infty}+\lambda_a f^2_{\infty}=0$, see e.g., \cite{Cai:2001dz,Hung:2011xb}.
Demanding that the quantity  $A$ is positive gives the following restriction on the coupling
\be\label{cond_coup_I}
\lambda_1< \f{R^2_A}{2(d-1)(d-2)}.
\ee

Using the couplings used in \cite{Hung:2011xb}, we can rewrite\footnote{where $\lambda_a$ here is same as $\lambda$ in \cite{Hung:2011xb}.} the coupling $\lambda_1=\f{2\lambda_a R^2_A}{(d-2)(d-3)}$,  in which case
\be\label{cond_coup_II}
\lambda_a< \f{d-3}{4(d-1)}.
\ee

The inclusion of the finite \lq{}tHooft coupling correction to the entanglement entropy functional does not  automatically make the entanglement entropy functional a minimum\footnote{It is suggested in \cite{Dong:2013qoa}  that when the extra piece other than the area of the co-dimension two surface term  in the entanglement entropy functional has the form of $f(R)$, where $R$ is the induced scalar curvature of the co-dimension two surface then one expects to have a minimum in the entanglement entropy functional.}. 

Upon demanding  the minimal condition on the entanglement entropy functional 
 puts a restriction on the coupling as written in eq(\ref{cond_coup_I}) and eq(\ref{cond_coup_II}).  Hence, we can interpret that  the minimality condition essentially says that the coupling has an upper bound which  is positive.   Moreover, the Jacobi test in the present case does not give anything interesting to leading linear order in $\lambda_1$, as the terms $DE,~E^2$ and $D^2$ in the $det(M)$ are quadratic order in $\lambda_1$.  

\paragraph{Discussion:} It is suggested in \cite{Fursaev:2013fta} that the strong   subadditivity  property\footnote{The strong subadditivity  property, $S(A)+S(B)\geq S(A\cup B)+S(A\cap B)$, is proven in the holographic case but without the higher derivative term in \cite{Headrick:2007km}. It is certainly interesting to ask whether the entanglement entropy functional as suggested, generically, in \cite{  Dong:2013qoa, Camps:2013zua} do automatically respect the strong subadditivity property  and the hypersurfce under study becomes a minimal surface. Moreover, we need to find the precise connection between the strong subadditivity and the minimal hypersurface.  }  should be obeyed by the entanglement entropy functional eq(\ref{assumption_EE}),  and the integration is done over a hypersurface which minimizes the entanglement entropy functional. We noticed that such minimality of the entanglement entropy functional  does not happen for all values of the couplings, $\lambda_a$, however, it does happen only when we put a serious restriction on the coupling $\lambda_a$ as in eq(\ref{cond_coup_II}). Hence, the imposition of the minimization condition  on the  entanglement entropy functional with the higher derivative term as suggested in \cite{Fursaev:2013fta} puts a  restriction on the coupling $\lambda_a$.

In $4+1$ dimensional AdS spacetime, it is suggested in  \cite{Buchel:2009tt} using the positivity of the energy fluxes and the causality that the Gauss-Bonnet (GB) coupling stays in a small window and can become a small negative number to a small positive number, which in our notation becomes $-\f{7}{36}\leq \lambda_a \leq \f{9}{100}$. From the study of the minimality of the entanglement entropy functional, we find  for $d=4$, that the coupling should have an upper bound, i.e., $\lambda_a< \f{1}{12}$. It is not known how to fix the  lower bound.

Generalizing it to arbitrary $d+1$ dimensional spacetime, it is found in \cite{Ge:2009eh,Buchel:2009sk,Camanho:2009hu} that the coupling, in our notation, should  stay in the following range
\be\label{cond_lambda_cft}
-\f{(d-2)(3d+2)}{4(d+2)^2}\leq \lambda_a \leq \f{(d-2)(d-3)(d^2-d+6)}{4(d^2-3d+6)^2}.
\ee
It is interesting to note that in the large $d$ limit, $d\ra\infty$, both eq(\ref{cond_coup_II}) and eq(\ref{cond_lambda_cft})  gives the same upper bound, namely, $1/4$.

The disagreement on the range of  the GB coupling suggests that the hypersurface under study does not necessarily minimizes the entanglement entropy\footnote{Let us note that the constraint on the coupling $\lambda_a$ follows (from eq(\ref{cond_coup_II})) by doing an analysis  only to leading order in the coupling. }. Hence, it remains an open question to know the precise nature of the hypersurface with higher derivative term in the 
entanglement entropy. 

\section{Conclusion}

The Ryu-Takayanagi (RT) conjecture gives an interesting proposal to calculate the entanglement entropy using a gravitational description. For a fixed time slice, the RT conjecture states that the entanglement entropy functional  is described by the area of a co-dimension two hypersurface. Moreover, the 
 co-dimension two hypersuface should be  determined in such a way that it minimizes the entanglement entropy functional.
In this paper, we have studied the consequences of  the  minimality condition  on the entanglement entropy functional, especially by performing the Legendre test and the Jacobi test. 
We have checked,  for the strip type entangling region, by studying  various  examples like thermal AdS solution, confining solution, hyperscale violating solution and the  black holes in the AdS spacetime  that it obeys necessarily the (weak)  minimality condition but not the sufficient condition. 

For our purpose, the outside of the black hole is described by the radial  coordinate that stays from the boundary $r=0$ to the horizon, $r=r_h$, whereas the inside is described by $r>  r_h$. Let us recall from the second variation of the area functional eq(\ref{2nd_var_area}) that it is the sign of the quantity, $A$, that  determines whether the area functional is  a minimum or a maximum. 
It is easy to notice   using the property of $g_{rr}$ as mentioned in section 2 and from eq(\ref{exp_A}),  that as long as we stay outside of the horizon, it gives a minimum. Once  we are inside the horizon, it gives a maximum. So, we may interpret,  it is the horizon that acts as a surface which separates the minimum area functional from the maximum.
 Hence, we can say that it is 
the RT conjecture that leads naturally to   the following conclusion:   we better stay outside of the horizon  if we want a minimum area. This finally allows us to conclude that the minimality of the area functional does not allow the co-dimension two hypersurface to enter into the black hole horizon. The same conclusion is reached\footnote{ In this case there does not exists any real valued solution of the embedding field, $X^M$, inside the horizon.} in \cite{Hubeny:2012ry} and more generally in \cite{Engelhardt:2013tra}.

In a recent study in \cite{Engelhardt:2013tra},  it is argued that  regions with negative extrinsic curvature cannot be accessed by any
hypersurfaces irrespective of whether it is of spacelike, timelike or null  type.
Let us recall that the imposition of the (weak) minimality condition gives us eq(\ref{min_area}), which is negative inside the  horizon.
 {\it A priori}, it is not clear whether there exists any  relationship\footnote{ One way to look at is as follows: the number of free indices that appear  in  eq(\ref{min_area}) is four whereas in the definition of the extrinsic curvature as in \cite{Engelhardt:2013tra}, it can be of maximum  three, for a hypersurface of co-dimension bigger than unity.} between the  extrinsic curvature studied in \cite{Engelhardt:2013tra} and   eq(\ref{min_area}). However, we do expect there should exist some kind of relation between these quantities because of the similarity in their behavior. In particular, for the black hole geometry, the quantity, $A$, as written in eq(\ref{exp_A}) shows  

\[ A^{-1} = \left\{
  \begin{array}{l l}
    +ve & \quad \textrm{for $r< r_h$ \quad (Outside the horizon)}\\
     ~ 0 & \quad \textrm{for $r= r_h$\quad    (On the horizon)}\\
    -ve & \quad \textrm{for $r> r_h$\quad    (Inside the horizon)}.
  \end{array} \right. \]

The extrinsic curvature shows precisely the similar type of behavior as reported in \cite{Engelhardt:2013tra}. The connection between these two quantities   are worth studying, which we leave for future studies. 
 
Moving onto the calculation of the entanglement entropy with higher derivative term, it is argued in  \cite{Fursaev:2013fta} that the hypersurface should be  minimal when the entanglement entropy  functional is described by eq(\ref{assumption_EE}). Upon applying such a minimality condition 
imposes  an important restriction on the Gauss-Bonnet coupling, $\lambda_1$. This is given in   eq(\ref{cond_coup_I}) and eq(\ref{cond_coup_II}), which  essentially gives an upper bound on the coupling. 
The bound so obtained does not match precisely with that derived in \cite{Buchel:2009sk,Camanho:2009hu}  using the positivity of the energy fluxes and the causality constraint. Hence, it is highly plausible that  theories with higher derivative term in the entanglement entropy functional does not have  hypersurfaces that are either minimal or maximal  in nature. So, the question about its nature remain to be seen in future studies.
 
\paragraph{Acknowledgement:}  It is a pleasure to thank the almighty for giving us enough energy to do this work. Certainly, no thanks to the Weather-God and his/her cabinet members for pushing the atmospheric temperature all the way to $45^{\circ}$ C during April-May and the humidity to 95\% in June-July. SSP would like to thank the computer center, IoP, Bhubaneswar.


\begin{thebibliography}{99}
\bibitem{Ryu:2006bv} 
S. Ryu and T. Takayanagi,  Phys. Rev. Lett. {\bf  96}, 181602 (2006), [arXiv:hep-th/0603001]; S. Ryu and T. Takayanagi, JHEP
0608, 045 (2006), [arXiv:hep-th/0605073].


\bibitem{Maldacena:1997re} 
  J.~M.~Maldacena,
  Adv.\ Theor.\ Math.\ Phys.\  {\bf 2}, 231 (1998)
  [hep-th/9711200];  S. S. Gubser, I. R. Klebanov
and A. M. Polyakov,  Phys. Lett.  {\bf B 428} (1998) 105, [arXiv:hep-th/9802109]; E. Witten, Adv. Theor. Math. Phys. {\bf 2}, 253 (1998), [arXiv:hep-th/9802150]; 
O. Aharony, S. S. Gubser, J. Maldacena, H. Ooguri and Y. Oz,  Phys. Rept, {\bf 323}, 183-386  (2000), [arXiv:hep-th/9905111].

\bibitem{Bianchi:2012ev} 
  E.~Bianchi and R.~C.~Myers,
 [arXiv:1212.5183 [hep-th]].


\bibitem{Hubeny:2007xt} 
  V.~E.~Hubeny, M.~Rangamani and T.~Takayanagi,
  JHEP {\bf 0707}, 062 (2007),
  [arXiv:0705.0016 [hep-th]].

\bibitem{Chen:2013qma} 
  B.~Chen and J.~-j.~Zhang,
  JHEP {\bf 07}, 185 (2013),
  [arXiv:1305.6767 [hep-th]].

\bibitem{Pal:2013fha} 
  S.~S.~Pal,
  Nucl.\ Phys.\ B {\bf 882}, 352 (2014),
  [arXiv:1312.0088 [hep-th]].
\bibitem{Erdmenger:2014tba} 
  J.~Erdmenger, M.~Flory and C.~Sleight,
  [arXiv:1401.5075 [hep-th]].
\bibitem{Bhattacharyya:2014yga} 
  A.~Bhattacharyya and M.~Sharma,
 [arXiv:1405.3511 [hep-th]].
\bibitem{Alishahiha:2013dca} 
  M.~Alishahiha, A.~F.~Astaneh and M.~R.~M.~Mozaffar,
  JHEP {\bf 1402}, 008 (2014),
  [arXiv:1311.4329 [hep-th]];
  M.~Alishahiha, A.~F.~Astaneh and M.~R.~M.~Mozaffar,
  Phys.\ Rev.\ D {\bf 89}, 065023 (2014),
  [arXiv:1310.4294 [hep-th]].


\bibitem{Myers:2012ed} 
  R.~C.~Myers and A.~Singh,
  JHEP {\bf 1204}, 122 (2012),
  [arXiv:1202.2068 [hep-th]].
 
\bibitem{Engelhardt:2013tra} 
  N.~Engelhardt and A.~C.~Wall,
  JHEP {\bf 1403}, 068 (2014),
  [arXiv:1312.3699 [hep-th]].
\bibitem{Hubeny:2012ry} 
  V.~E.~Hubeny,
  JHEP {\bf 1207}, 093 (2012),
  [arXiv:1203.1044 [hep-th]].
 %
 %
\bibitem{Fursaev:2013fta} 
  D.~V.~Fursaev, A.~Patrushev and S.~N.~Solodukhin,
  Phys.\ Rev.\ D {\bf 88},  044054 (2013),
  [arXiv:1306.4000 [hep-th]].
 \bibitem{Dong:2013qoa} 
  X.~Dong,
  JHEP {\bf 1401}, 044 (2014),
  [arXiv:1310.5713 [hep-th]].
\bibitem{Camps:2013zua} 
  J.~Camps,
  JHEP {\bf 1403}, 070 (2014),
  [arXiv:1310.6659 [hep-th]].
\bibitem{Fursaev:2006ih} 
  D.~V.~Fursaev,
  JHEP {\bf 0609}, 018 (2006),
  [hep-th/0606184].
\bibitem{deBoer:2011wk} 
  J.~de Boer, M.~Kulaxizi and A.~Parnachev,
  JHEP {\bf 1107}, 109 (2011),
  [arXiv:1101.5781 [hep-th]].
\bibitem{Hung:2011xb} 
  L.~-Y.~Hung, R.~C.~Myers and M.~Smolkin,
  JHEP {\bf 1104}, 025 (2011),
  [arXiv:1101.5813 [hep-th]].
\bibitem{Buchel:2009sk} 
  A.~Buchel, J.~Escobedo, R.~C.~Myers, M.~F.~Paulos, {\it et. al.},
  JHEP {\bf 1003}, 111 (2010),
  [arXiv:0911.4257 [hep-th]].

\bibitem{Camanho:2009hu} 
  X.~O.~Camanho and J.~D.~Edelstein,
  JHEP {\bf 1006}, 099 (2010).
  [arXiv:0912.1944 [hep-th]].
\bibitem{Pal:2012zn} 
  S.~S.~Pal,
  JHEP {\bf 1304}, 007 (2013),
  [arXiv:1209.3559 [hep-th]].


\bibitem{Cai:2001dz} 
  R.~-G.~Cai,
  Phys.\ Rev.\ D {\bf 65}, 084014 (2002),
  [hep-th/0109133].

\bibitem{Nojiri:2001aj} 
  S.~'i.~Nojiri and S.~D.~Odintsov,
  Phys.\ Lett.\ B {\bf 521}, 87 (2001),
  [Erratum-ibid.\ B {\bf 542}, 301 (2002)]
  [hep-th/0109122];
  M.~Cvetic, S.~'i.~Nojiri and S.~D.~Odintsov,
  Nucl.\ Phys.\ B {\bf 628}, 295 (2002),
  [hep-th/0112045].


\bibitem{Cho:2002hq} 
  Y.~M.~Cho and I.~P.~Neupane,
  Phys.\ Rev.\ D {\bf 66}, 024044 (2002),
  [hep-th/0202140];
I.~P.~Neupane,
  Phys.\ Rev.\ D {\bf 67}, 061501 (2003),
  [hep-th/0212092];
I.~P.~Neupane,
  Phys.\ Rev.\ D {\bf 69}, 084011 (2004),
  [hep-th/0302132].














\bibitem{Buchel:2009tt} 
  A.~Buchel and R.~C.~Myers,
  JHEP {\bf 0908}, 016 (2009),
  [arXiv:0906.2922 [hep-th]].




\bibitem{Headrick:2007km} 
  M.~Headrick and T.~Takayanagi,
  Phys.\ Rev.\ D {\bf 76}, 106013 (2007),
  [arXiv:0704.3719 [hep-th]].
\bibitem{Ge:2009eh} 
  X.~-H.~Ge and S.~-J.~Sin,
  JHEP {\bf 0905}, 051 (2009),
  [arXiv:0903.2527 [hep-th]];
  X.-H.~Ge, S.-J.~Sin, S.-F.~Wu and G.-H.~Yang,
  Phys.\ Rev.\ D {\bf 80}, 104019 (2009),
  [arXiv:0905.2675 [hep-th]]; X.-H.~Ge, Y.~Matsuo, F.-W.~Shu, S.-J.~Sin and T.~Tsukioka,
  JHEP {\bf 0810}, 009 (2008),
  [arXiv:0808.2354 [hep-th]].




\bibitem{Jacobson:1993xs} 
  T.~Jacobson and R.~C.~Myers,
  Phys.\ Rev.\ Lett.\  {\bf 70}, 3684 (1993),
  [hep-th/9305016].



\bibitem{Klebanov:2007ws} 
  I.~R.~Klebanov, D.~Kutasov and A.~Murugan,
  Nucl.\ Phys.\ B {\bf 796}, 274 (2008),
  [arXiv:0709.2140 [hep-th]].


 

\end{thebibliography}
\end{document}